\documentclass[prd,aps,showpacs,preprintnumbers,amssymb]{revtex4}
\usepackage{graphicx}

\usepackage{verbatim}
\usepackage{xcolor}
\begin{document}

\title{%
\hfill{\normalsize\vbox{%
 }}\\
{Naturally perturbed S$_3$ neutrinos}}

\author{Renata Jora
$^{\it \bf a}$~\footnote[1]{Email:
  rjora@theory.nipne.ro }}

\author{Joseph Schechter
 $^{\it \bf b}$~\footnote[2]{Email:
 schechte@phy.syr.edu}}

\author{M. Naeem Shahid
$^{\it \bf b,c}$~\footnote[3]{Email:
   mnshahid@phy.syr.edu            }}

\affiliation{$^{\bf a}$ National Institute of Physics and Nuclear Engineering, PO Box MG-6, Bucharest-Magurele, Romania}

\affiliation{$^ {\bf \it b}$ Department of Physics,
 Syracuse University, Syracuse, NY 13244-1130, USA}
 
 \affiliation{$^ {\bf \it c}$National Centre for Physics, Quaid-i-Azam University Campus, 45320 Islamabad, Pakistan}

\date{\today}

\begin{abstract}
We simplify and extend our previous model for the masses and mixing matrix of three Majorana neutrinos based on permutation symmetry $S_3$ and the perturbations which violate this symmetry. The perturbations are arranged such that we get the smaller solar neutrinos mass difference at second order. We work out the corrections to the tribimaximal mixing matrix with the non-zero value for $s_{13}$ and the conventional CP-violating phase. It is shown that the results of the model are comparable with the global analysis of neutrino oscillation data.
\end{abstract}


\maketitle

\section{Introduction}
In the present paper, a certain approach  to understanding
    neutrino mixing, proposed in \cite{jns} and further studied in \cite{cw}, \cite{jss2009}, \cite{jss2010}, \cite{dgg1} and \cite{dgg2}
    will be examined in more detail. This approach is based on assuming the permutation group $S_3$ symmetry for the lepton sector to be a starting point. The three neutrinos are arranged to be degenerate in this limit. Analogously to the SU(3) flavor and isospin breaking in QCD we add two breaking terms in different $S_3$ ``directions" with different strengths. Of course there have been many related treatments of this topic in the literature \cite{w} - \cite{ks}.
    
    There are two new features compared to our earlier papers. First the order of the two perturbations will be arranged to correspond to the observed hierarchy between the ``atmospheric" neutrino mixing and the ``solar" neutrino mixing. Secondly the effect of ``Dirac type" CP-violation will be included in addition to the previous Majorana type CP-violation.

\section{Brief review and new changes}

In \cite{jns} the ``unperturbed" neutrino mass matrix was taken to be the $S_3$ symmetric form, 
\begin{equation}
M^0_\nu=\alpha
\left[
\begin{array}{ccc}
1&0&0\\
0&1&0\\
0&0&1
\end{array}
\right]+\beta
\left[
\begin{array}{ccc}
1&1&1\\
1&1&1\\
1&1&1\\
\end{array}
\right] \equiv \alpha {\bf 1}+\beta d .
\label{solution}
\end{equation}
$\alpha$ and $\beta$ are, in general, complex numbers
for the case of Majorana neutrinos
 while $d$ is
usually called the ``democratic" matrix.

This is diagonalized by,
\begin{equation}
R^T(\alpha{\bf 1}+{\beta}d)R=\left[
\begin{array}{ccc}
\alpha&0&0 \\
0&\alpha+3\beta&0 \\
0&0&\alpha \\
\end{array}
\right],
\label{complexeigenvalues}
\end{equation}
where $R$ has the so called ``tribimaximal" form,
\begin{equation}
    R=\left(%
\begin{array}{ccc}
  \frac{-2}{\sqrt{6}} & \frac{1}{\sqrt{3}} & 0 \\
  \frac{1}{\sqrt{6}} & \frac{1}{\sqrt{3}} & \frac{1}{\sqrt{2}} \\
  \frac{1}{\sqrt{6}} & \frac{1}{\sqrt{3}} & \frac{-1}{\sqrt{2}} \\
\end{array}%
\right).
\label{R}
\end{equation}

\begin{figure}[htbp]
\centering
\rotatebox{0}
{\includegraphics[width=10cm,height=10cm,clip=true]{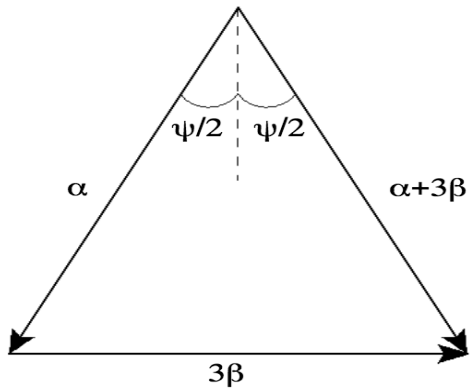}}
\caption[]{Isosceles triangle with angle $\psi$ between
the equal length 2-vectors $\alpha$ and $\alpha + 3\beta$.}
\label{triangle}
\end{figure}

To insure the initial neutrino degeneracy, the complex numbers $\alpha$ and $\alpha + 3 \beta $ are arranged as shown in Fig. (\ref{triangle}). We introduce the notation, 
\begin{equation}
\alpha \equiv -i|\alpha|e^{-i\psi/2}.
\label{alpha}
\end{equation}
The physical phase $\psi$ lies in the range:
\begin{equation}
o<\psi\leq \pi,
\label{psirange}
\end{equation}
and $|\alpha|$ is related to $\beta$ by,
\begin{equation}
|\alpha|=\frac{3\beta}{2sin(\psi/2)}.
\label{alphabeta}
\end{equation}

In \cite{jss2009} the effect of a perturbation matrix of the form,
\begin{equation}
\Delta=\left(%
\begin{array}{ccc}
  0 & 0 & 0 \\
  0 & t & u \\
  0 & u & t \\
\end{array}%
\right),
\label{23pert}
\end{equation}
was considered with $t$ and $u$ real.

In the present work, we will first reduce the number of parameters by setting $u= -t$. This has the effect of eliminating the smaller ``solar neutrino" mass splitting at the first order of perturbation. That, as well as CP-violation effects will be introduced at the next order of perturbation via the matrix,

\begin{equation}
\Delta^{\prime}=ve^{-i\phi}\left(%
\begin{array}{ccc}
  1 & -1 & 0 \\
  -1 & 1 & 0 \\
  0 & 0 & 0 \\
\end{array}%
\right).
\label{12pert}
\end{equation}

Here $v$ is real and $\phi$ is a CP-violating phase. This represents a variation of the similar  matrix $\Delta^\prime$ used in the doubly perturbed model of \cite{jss2010}.

Hence the present model has only the three perturbation parameters $t$, $v$ and $\phi$. $t$ will be considered to be of first order strength while $v$ will be taken to be second order. On the other hand the CP phase will not necessarily be taken small. The $ve^{-i\phi}$ term will introduce both non-zero solar neutrino masses and Dirac type CP-violation.  The main work will be to find the eigenvalues and eigenvectors of,
\begin{equation}
M_\nu=\alpha{\bf 1} +\beta d +\Delta +\Delta^\prime.
\label{mainmatrix}
\end{equation}

\section{Diagonalizing the neutrino mass matrix}
Since we are working in a basis where the zeroth order
piece is diagonalized by the tribimaximal matrix $R$,
we must bring to diagonal form the matrix:
\begin{eqnarray}
 R^TM_\nu R =
\left(%
\begin{array}{ccc}
  -i|\alpha|e^{-i\psi/2}+\frac{3}{2}ve^{-i\phi} & 0 & \frac{\sqrt{3}}{2}ve^{-i\phi}\\
 0  &  -i|\alpha|e^{-i\psi/2}+3\beta &0 \\
   \frac{\sqrt{3}}{2}ve^{-i\phi} & 0 & -i|\alpha|e^{-i\psi/2}+2t+\frac{1}{2}ve^{-i\phi}\\
\end{array}
\right).
\label{tobediag}
\end{eqnarray}
Denoting the non-trivial 2 $\times$ 2, (13) sub-block by $X$,
\begin{eqnarray}
 X =
\left(%
\begin{array}{cc}
  -i|\alpha|e^{-i\psi/2}+\frac{3}{2}ve^{-i\phi} & \frac{\sqrt{3}}{2}ve^{-i\phi}\\
   \frac{\sqrt{3}}{2}ve^{-i\phi}  & -i|\alpha|e^{-i\psi/2}+2t+\frac{1}{2}ve^{-i\phi}\\
\end{array}
\right),
\label{13block}
\end{eqnarray}
 we diagonalize the hermitian quantity $X^\dagger X$ with a unitary matrix $U$ in the usual way,
\begin{equation}
U^\dagger X^\dagger X U ={\rm Real, diagonal}.
\end{equation} 
This implies, as actually required to bring the Majorana type neutrino mass matrix to diagonal form,
\begin{equation}
U^T X U ={\rm diagonal},
\end{equation} 
Explicitly,
\begin{equation}
U \approx
\left(%
\begin{array}{cc}
  1  & \epsilon\\
  -\epsilon^*  &1 \\
\end{array}%
\right),
\label{U}
\end{equation}
where to second order,
\begin{equation}
\epsilon =\frac{\sqrt{3}v sin(\frac{\psi}{2}-\phi)}{4t sin(\frac{\psi}{2})}\left[1+\frac{t}{|\alpha| sin(\frac{\psi}{2})}-\frac{t}{|\alpha| sin(\frac{\psi}{2}-\phi)}e^{i\phi}\right].
\label{eps}
\end{equation}
Here we used the approximations that $t$ is first order, $v$ is second order and $v/t$ is, for consistency, first order. We trivially upgrade this to the 3 $\times$ 3 level as,
\begin{equation}
Z=
\left(%
\begin{array}{ccc}
  1 & 0 & \epsilon \\
    0 & 1 & 0 \\
  -\epsilon^*&0&1  \\
\end{array}%
\right),
\label{Z}
\end{equation}
satisfying,
\begin{equation}
Z^T R^T M_\nu RZ =diag(d_1,d_2, d_3),
\label{complexdiag}
\end{equation}
with,
\begin{eqnarray}
d_1&=&-i|\alpha|e^{-i\psi/2}+\frac{3}{2}ve^{-i\phi},
\nonumber \\
d_2&=& -i|\alpha|e^{-i\psi/2}+3\beta,
\nonumber \\
d_3&=& -i|\alpha|e^{-i\psi/2}+2t+\frac{1}{2}ve^{-i\phi}.
\label{evs}
\end{eqnarray}
The neutrino masses $m_i$ are the magnitudes of the $d_i$:
\begin{eqnarray}
m_1&\approx& |\alpha|- \frac{3v}{2} sin(\frac{\psi}{2}-\phi),
\nonumber \\
m_2&\approx& |\alpha|,
\nonumber \\
m_3 &\approx&|\alpha| -2t sin(\frac{\psi}{2})-\frac{v}{2} sin(\frac{\psi}{2}-\phi).
\label{numasses}
\end{eqnarray}

These mass parameters were made real, positive
 by the introduction
of the phase matrix:
\begin{equation}
P=
\left(%
\begin{array}{ccc}
  e^{-i\tau} & 0 & 0  \\
  0 & e^{-i\sigma} & 0 \\
  0 & 0 & e^{-i\rho} \\
\end{array}%
\right),
\label{phasematrix}
\end{equation}
where,
\begin{eqnarray}
\tau&\approx&  \frac{\pi}{2}+\frac{1}{2}tan^{-1}\left[cot(\frac{\psi}{2})\left(1+\frac{3vcos(\frac{\psi}{2}-\phi)}{|\alpha|sin(\psi)}\right) \right],
\nonumber \\
\sigma&\approx& \pi-\frac{1}{2}tan^{-1}\left[cot(\frac{\psi}{2})\right]=\frac{3\pi}{2}+\frac{\psi}{2},
\nonumber \\
\rho&\approx& \frac{\pi}{2}+\frac{1}{2}tan^{-1}\left[cot(\frac{\psi}{2})\left(1+\frac{2t}{|\alpha|sin(\frac{\psi}{2})}+\frac{vcos(\frac{\psi}{2}-\phi)}{|\alpha|sin(\psi)}\right) \right].
\label{majphases}
\end{eqnarray}

The entire approximate diagonalization may be presented as,
\begin{equation}
K^T(\alpha {\bf 1}+\beta d +\Delta+\Delta^\prime)K =
\left(%
\begin{array}{ccc}
  m_1 & 0 & 0  \\
  0 & m_2 & 0 \\
  0 & 0 & m_3 \\
\end{array}%
\right),
\label{entirediag}
\end{equation}
where,
\begin{equation}
K=RZP
\label{fullmixingmat}
\end{equation}
is the full neutrino mixing matrix.
Explicitly,
\begin{equation}
    K=\left(%
\begin{array}{ccc}
  \frac{-2}{\sqrt{6}} & \frac{1}{\sqrt{3}} & \frac{-2\epsilon}{\sqrt{6}} \\
  \frac{1}{\sqrt{6}}-\frac{\epsilon^*}{\sqrt{2}} & \frac{1}{\sqrt{3}} & \frac{1}{\sqrt{2}}+\frac{\epsilon}{\sqrt{6}} \\
  \frac{1}{\sqrt{6}}+\frac{\epsilon^*}{\sqrt{2}} & \frac{1}{\sqrt{3}} & \frac{-1}{\sqrt{2}} +\frac{\epsilon}{\sqrt{6}}\\
\end{array}%
\right)P,
\label{K}
\end{equation}
where $\epsilon$ is given in Eq. (\ref{eps}).

If one assumes that the factor matrix associated with diagonalizing the charged lepton mass matrix contains only small angles, Eq. (\ref{K}) should provide a reasonable approximation to the leptonic analog of the Cabibbo-Kobayashi-Maskawa mixing matrix. Such an assumption seems plausible since, for example, in Grand Unified Models \cite{bkns1} and \cite{bkns2}, the charged lepton contributions are similar in magnitude to the quark contributions (which do involve small angles). In the present $S_3$ framework, it was shown in section V of \cite{jns} that the charged lepton contribution can be arranged to vanish. Clearly it is sensible to first analyze the ``neutrino dominance" case in detail.

\section{Parameter evaluation from experiment}

First we have important
 information from neutrino oscillation
 experiments \cite{superK}-\cite{minos}.
  It is known that \cite{A}
\begin{eqnarray}
A&\equiv& m_2^2-m_1^2= (7.50 \pm 0.20)\times 10^{-5} {\rm eV}^2,
\nonumber \\
B&\equiv& |m_3^2-m_2^2|= (2.32^{ +0.12}_{-0.08})\times 10^{-3} {\rm eV}^2.
\label{AB}
\end{eqnarray}
Also, constraints on cosmological structure formation yield
\cite{cosmobound1} - \cite{cosmobound2} a
rough bound,
\begin{equation}
m_1+m_2+m_3 \leq 0.3 {\rm eV} .
\label{cosmo}
\end{equation}

The two allowed spectrum types are:
\begin{eqnarray}
&&{\rm Type 1  \,(normal\,hierarchy)}:\hspace{1cm} m_3>m_2>m_1,
\nonumber \\
&&{\rm Type 2 \, (inverted\, hierarchy)}:\hspace{1cm} m_2>m_1>m_3.
\label{spectrumtype}
\end{eqnarray}.

From Eqs. (\ref{numasses}) we write:
\begin{eqnarray}
 m_2^2-m_1^2&=&3|\alpha| v sin(\frac{\psi}{2}-\phi) ,
\nonumber \\
 m_3^2-m_2^2&=&-4|\alpha| t sin(\frac{\psi}{2})-|\alpha| v sin(\frac{\psi}{2}-\phi).
\label{deltanus}
\end{eqnarray}
For simplicity we will not try to determine all the parameters together. Rather we bound $|\alpha|$ at zeroth order, evaluate $t|\alpha|$ at first order and evaluate $v|\alpha|$ at second order.
 
\subsection{Zeroth order}
At this order the three neutrino masses are degenerate. With $v=t=0$ Eqs. (\ref{numasses}) and (\ref{cosmo}) imply the bound,  
\begin{equation}
|\alpha|\lesssim 0.1 {\rm eV}.
\label{cosmozero}
\end{equation}

\subsection{First order}
At this order the degeneracy between $m_1$ and $m_2$ is still preserved. With $v=0,t\neq 0$ one has,
\begin{eqnarray}
m_{2}^2 - m_{1}^2 &=& 0 \nonumber\\
m_{3}^2 - m_{2}^2 &=& -4 |\alpha |tsin(\frac{\psi}{2}),
\end{eqnarray}
which implies,
\begin{equation}
t|\alpha|= \frac{m_3^2-m_2^2}{-4 sin(\frac{\psi}{2})}.
\label{alphat}
\end{equation}
This yields Table I for $t|\alpha|$ for several values of the Majorana-type phase $\psi$ and choice of the neutrino mass ``hierarchy".
\begin{table}[htbp]
\begin{center}
\begin{tabular}{c|c|c}
\hline
  $\psi$ & $t|\alpha|$ for normal hierarchy (eV$^2$) &$t|\alpha|$ for inverted hierarchy (eV$^2$) \\
\hline \hline
$\pi$ & $-$5.8 $\times$ 10$^{-4}$ &  5.8 $\times$ 10$^{-4}$       \\
\hline
$\pi/2$ & $ -$8.2 $\times$ 10$^{-4}$  & 8.2 $\times$ 10$^{-4}$      \\
\hline
$\pi/4$ &  $-$15.2 $\times$ 10$^{-4}$  & 15.2 $\times$ 10$^{-4}$   \\
\hline
\end{tabular}
\end{center}
\caption[]{$t|\alpha|$ for different values of $\psi$.}
 \label{ttab}
\end{table}

\subsection{Second order}
At second order the degeneracy between $m_1$ and $m_2$ is broken and $v,t\neq 0$. Using Eqs. (\ref{deltanus}) this determines $v|\alpha|$,
\begin{equation}
v|\alpha| =\frac{ m_2^2-m_1^2}{3sin(\frac{\psi}{2}-\phi)}.
\label{alphav}
\end{equation}

We can plot $v|\alpha|$ vs $\phi$ for various values of $\psi$. As an example, the plot for $\psi = \pi$ is given in Fig. 2.

 \begin{figure}[htbp]
\centering
\rotatebox{0}
{\includegraphics[]{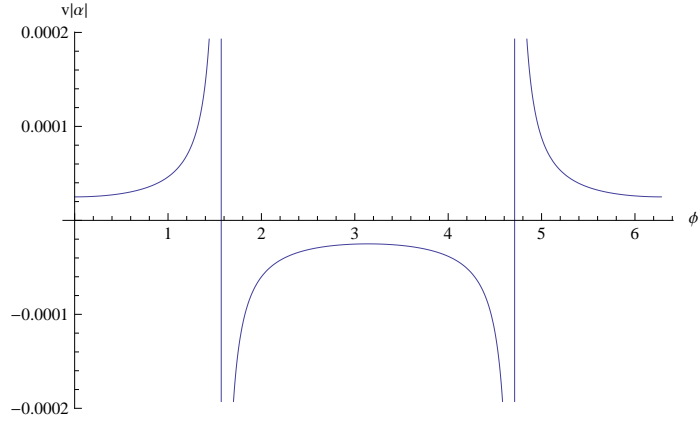}}
\caption[]{Plot of $v|\alpha|$ versus $\phi$ for the choice $\psi =\pi$}
\label{valphavsphi}
\end{figure}

The plots for other values of $\psi$ are similar. A more detailed comparison is furnished in Table II.

\begin{table}[htbp]
\begin{center}
\begin{tabular}{c|c|c|c|c|c|c|c}
\hline
  $\psi$ & $v|\alpha |$ for $\phi =0$& $v|\alpha |$ for $\phi =\pi/3$ & $v|\alpha |$ for $\phi =2\pi/3$ & $v|\alpha |$ for $\phi = \pi$ & $v|\alpha |$ for $\phi = 4\pi/3$ & $v|\alpha |$ for $\phi = 5\pi/3$  \\
\hline \hline
$\pi$ & 2.5 $\times$ 10$^{-5}$ & 5.0 $\times$ 10$^{-5}$ & $-$ 5.0 $\times$ 10$^{-5}$ & $-$ 2.5 $\times$ 10$^{-5}$ & $-$ 5.0 $\times$ 10$^{-5}$ & 5.0 $\times$ 10$^{-5}$ \\
\hline
$\pi/2$ & 3.5 $\times$ 10$^{-5}$ & $-$9.6 $\times$ 10$^{-5}$ & $-$ 2.6 $\times$ 10$^{-5}$ & $-$ 3.5 $\times$ 10$^{-5}$ & 9.6 $\times$ 10$^{-5}$ & 2.6 $\times$ 10$^{-5}$ \\
\hline
$\pi/4$ & 6.5 $\times$ 10$^{-5}$ & $-$ 4.1 $\times$ 10$^{-5}$ & $-$ 2.5 $\times$ 10$^{-5}$ & $-$ 6.5 $\times$ 10$^{-5}$ & 4.1 $\times$ 10$^{-5}$ & 2.5 $\times$ 10$^{-5}$ \\
\hline
\end{tabular}
\end{center}
\caption[]{$v|\alpha|$ (in eV$^2$) for different values of $\psi$ and $\phi$.}
 \label{vtab}
\end{table}

Note that the ratio $v/t$ of second order to first order parameters depends somewhat on the CP phase, $\phi$ but is less than 1/10 in magnitude.

\subsection{The parameter $\epsilon$ and $s_{13}$}
The complex mixing parameter $\epsilon$ in Eq. (\ref{eps}) can be written as,
\begin{equation}
\epsilon=| \epsilon |e^{-i\delta},
\label{epsrewrite}
\end{equation}
where,
\begin{eqnarray}
| \epsilon | &\approx& \frac{\sqrt{3}v sin(\frac{\psi}{2}-\phi)}{4t sin(\frac{\psi}{2})}\left[1+\frac{t}{|\alpha| sin(\frac{\psi}{2})}-\frac{tcos\phi}{|\alpha| sin(\frac{\psi}{2}-\phi)}\right]\nonumber\\
&=&\frac{\sqrt{3}vsin(\frac{\psi}{2}-\phi)}{4tsin(\frac{\psi}{2})}\left[1-\frac{tcos(\frac{\psi}{2})sin\phi}{|\alpha|sin(\frac{\psi}{2})sin(\frac{\psi}{2}-\phi)}\right].
\label{epsmod}
\end{eqnarray}
The phase $\delta$ is given as,
\begin{equation}
tan\delta \approx  \frac{tsin\phi}{|\alpha| sin(\frac{\psi}{2}-\phi)}.
\label{epsphase}
\end{equation}
As mentioned above, we are keeping terms linear in $t/|\alpha|$, $v/t$ and $v/|\alpha|$. Using this approximation we get,
\begin{equation}
sin\delta \approx \frac{tsin\phi}{|\alpha| sin(\frac{\psi}{2}-\phi)},\quad cos\delta \approx 1.
\label{epsphase1}
\end{equation}

Writing down elements of the matrix in Eq. (\ref{K}) in terms of their magnitudes and phases yields, 

\begin{equation}
K=\left(\begin{array}{ccc}
\frac{-2}{\sqrt{6}} & \frac{1}{\sqrt{3}} & \frac{-2}{\sqrt{6}}|\epsilon|e^{-i\delta}\\
\left(\frac{1}{\sqrt{6}}-\frac{|\epsilon|}{\sqrt{2}}\right)e^{3i\omega} & \frac{1}{\sqrt{3}} & \left(\frac{1}{\sqrt{2}}+\frac{|\epsilon|}{\sqrt{6}}\right)e^{i\omega}\\
\left(\frac{1}{\sqrt{6}}+\frac{|\epsilon|}{\sqrt{2}}\right)e^{-3i\omega} & \frac{1}{\sqrt{3}} & \left(\frac{-1}{\sqrt{2}}+\frac{|\epsilon|}{\sqrt{6}}\right)e^{-i\omega}
\end{array}\right)P,
\label{kexpand}
\end{equation}
where,

\begin{equation}
\omega=-tan^{-1}\left(\frac{|\epsilon|sin\delta}{\sqrt{3}}\right)\approx-\frac{|\epsilon|sin\delta}{\sqrt{3}}\approx - \frac{v}{4|\alpha |}\frac{sin\phi}{sin\frac{\psi}{2}}.
\label{phases}
\end{equation}
We would like to identify the transformation matrix $K$ with a conventional form of the type:

\begin{equation}
K_{exp}P=\left(\begin{array}{ccc}
c_{12}c_{13} & s_{12}c_{13} & s_{13}e^{-i\gamma}\\
-s_{12}c_{23}-c_{12}s_{13}s_{23}e^{i\gamma} & c_{12}c_{23}-s_{12}s_{13}s_{23}e^{i\gamma} & c_{13}s_{23}\\
s_{12}s_{23}-c_{12}s_{13}c_{23}e^{i\gamma} & -c_{12}s_{23}-s_{12}s_{13}c_{23}e^{i\gamma} & c_{13}c_{23}
\end{array}\right)P.
\label{symparam}
\end{equation}
See, for example, Eq (25) of \cite{jss2010}. 

If we look at the matrix in Eq. (\ref{kexpand}), it seems like there are some extra phases, for example, the elements (23) and (33) have non-zero phases while (22) and (32) are without phases. In order to compare it with the above parametrization, we should take out these phases to bring it to the desired form. Let us multiply the matrix $K$ in Eq. (\ref{kexpand}) by the diagonal matrices of phases $L$ and $Q$ from left and right. The resultant matrix will be identified with $K_{exp}$,

\begin{equation}
LKQ=K_{exp}.
\label{phasetrans}
\end{equation}
 
\begin{equation}
LKQ=\left(\begin{array}{ccc}
\frac{2}{\sqrt{6}} & -\frac{1}{\sqrt{3}} & -\frac{2|\epsilon|}{\sqrt{6}}e^{-i\delta}\\
\left(\frac{1}{\sqrt{6}}-\frac{|\epsilon|}{\sqrt{2}}\right)e^{2i\omega} & \frac{e^{-i\omega}}{\sqrt{3}} & -\frac{1}{\sqrt{2}}-\frac{|\epsilon|}{\sqrt{6}}\\
\left(\frac{1}{\sqrt{6}}+\frac{|\epsilon|}{\sqrt{2}}\right)e^{-2i\omega} & \frac{e^{i\omega}}{\sqrt{3}} & \frac{1}{\sqrt{2}}-\frac{|\epsilon|}{\sqrt{6}}
\end{array}\right)P^\prime
\label{standardform}
\end{equation}
Here,
\begin{equation}
L=\left(\begin{array}{ccc}
-1 & 0 & 0\\
0 & e^{-i\omega} & 0\\
0 & 0 & e^{i\omega}
\end{array}\right),\quad Q=\left(\begin{array}{ccc}
1 & 0 & 0\\
0 & 1 & 0\\
0 & 0 & -1
\end{array}\right).
\label{extraphases}
\end{equation}
 Note that the mixing matrix obtained in Eq. (\ref{standardform}) is a phase rotation of the triminimal mixing matrix previously considered by some authors \cite{trimax}.

Now the phase matrix $L$ gets absorbed due to arbitrary phases for the charged leptons which will ``sit" on the left side. What remains is the standard form when we identify $P^\prime =PQ$ as the Majorana phases. This enables one to easily compare with the three conventional CP conserving angles and the conventional Dirac phase.

Comparing the matrices in Eqs. (\ref{symparam}) and (\ref{standardform}), we first note that $\epsilon$ in Eq. (\ref{K}) is related to $s_{13}$,

\begin{equation}
s_{13}=\frac{-2|\epsilon|}{\sqrt{6}},\quad \delta=\gamma.
\label{epss13}
\end{equation}

Also $s_{23}$ and $s_{12}$ can be read off:

\begin{equation}
s_{23}=-\frac{1}{\sqrt{2}}-\frac{|\epsilon|}{\sqrt{6}},\quad s_{12}=-\frac{1}{\sqrt{3}}=-0.58.
\label{k23}
\end{equation}

Expanding $s_{23}$ around its ``tribimaximal value" as $s_{23} =[s_{23}]_{TBM} + \delta s_{23}$, one gets:

\begin{equation}
(s_{23})^2 \approx \frac{1}{2} + \sqrt{2} \delta s_{23},
\label{s23expand}
\end{equation}

where $\delta s_{23}$ measures the deviation of $s_{23}$ from its tribimaximal value.

From Eqs. (\ref{epss13}) and (\ref{k23}), we find the interesting relation,
\begin{equation}
s_{13} =-2 \delta s_{23}.
\label{s13s23rel}
\end{equation}
We should point out here that some similar results were obtained in one of our earlier papers (see Eq. (27) of \cite{jss2010}), without the CP phase and with additional parameters. Many authors have also got the same result, see for example \cite{dgr1} - \cite{dgr4} .

Using the results of the global 3$\nu$ oscillation analysis \cite{fogli1} - \cite{fogli2}, the numerical values of $|\epsilon|$ and $s_{13}$ are illustrated in Table \ref{eps2tab} for both normal (left table) and inverted (right table) hierarchies. It is encouraging that these values for $|s_{13}|$ are very well	in agreement with the analysis of neutrino oscillation data. The best fit value for $|s_{13}|$ is 0.157 for normal hierarchy and 0.158 for inverted hierarchy which are also close to the predicted values at $1\sigma$, $2\sigma$ and $3\sigma$ levels.

Note also that the magnitude of  parameter $\epsilon$ and $s_{13}$ are only determined from the deviation in the mixing angle $s_{23}$ so the corrections to $s_{23}$ are fixed as our input. The mixing angle $s_{12}$ does not get any correction in this model but it is close to the best fit value of 0.56 from the data.

\begin{table}[htbp]
\begin{center}
\begin{tabular}{c|c|c|c}
   &  $1\sigma$ & $2\sigma$ & $3\sigma$   \\
\hline \hline
$(|\delta s_{23}|)_{data}$ &  0.061 & 0.10 & 0.11 \\
\hline
$|\epsilon|$ & 0.149 & 0.24 & 0.27 \\
\hline
$|s_{13}|$ & 0.122 & 0.20 & 0.22 \\
\hline
$(s_{13})_{data}$ & 0.157 & 0.16 & 0.16 \\
\hline
\end{tabular}
\quad
\begin{tabular}{c|c|c|c}
   &  $1\sigma$ & $2\sigma$ & $3\sigma$   \\
\hline \hline
$(|\delta s_{23}|)_{data}$ &  0.069 & 0.09 & 0.11 \\
\hline
$|\epsilon|$ & 0.169 & 0.22 & 0.27 \\
\hline
$|s_{13}|$ & 0.138 & 0.18 & 0.22 \\
\hline
$(s_{13})_{data}$ & 0.158 & 0.16 & 0.16 \\
\hline
\end{tabular}
\end{center}
\caption[]{Numerical values of $|\epsilon|$ and $s_{13}$ for normal (left) and inverted (right) hierarchies.}
 \label{eps2tab}
\end{table}

Finally the phase $\delta$ in Eq. (\ref{epsphase}) can be estimated using Table \ref{ttab}. Some values are given in Table \ref{deltatab} for both mass hierarchies. The best fit value for the phase $\delta$ in the global analysis of neutrinos is $0.80\pi$ for normal hierarchy and $-0.03\pi$ for inverted hierarchy. It can be seen that our values are close to the values for inverted hierarchy but an order smaller for normal hierarchy. 
\begin{table}[htbp]
\begin{center}
\begin{tabular}{c|c|c|c|c|c|c|c}
\hline
  $\psi$ & $\delta$ for $\phi =0$& $\delta$ for $\phi =\pi/3$ & $\delta$ for $\phi =2\pi/3$ & $\delta$ for $\phi = \pi$ & $\delta$ for $\phi = 4\pi/3$ & $\delta$ for $\phi = 5\pi/3$  \\
\hline \hline
$\pi$ & 0 & $-$ 0.03$\pi$ & 0.03$\pi$ & 0 & $-$ 0.03$\pi$ & 0.03$\pi$ \\
\hline
$\pi /2$ & 0 & 0.09$\pi$ & 0.02$\pi$ & 0 & 0.09$\pi$ & 0.02$\pi$ \\
\hline
$\pi /4$ & 0 & 0.07$\pi$ &  0.04$\pi$ & 0 &  0.07$\pi$ & 0.04$\pi$ \\
\hline
\end{tabular}
\end{center}
\caption[]{$\delta$ for different values of $\psi$ and $\phi$ for normal hierarchy (opposite sign for inverse hierarchy).}
 \label{deltatab}
\end{table}

\section{Summary and discussion}

In this work, we studied the model \cite{jns}, \cite{jss2009} and \cite{jss2010} for neutrino masses and mixing based on the permutation symmetry $S_3$ in more detailed and simplified way. In Sec. II, we presented the outlines and differences from previously studied models. Our starting point was the full $S_3$ invariance, with parameter $|\alpha|$, where we imposed the unperturbed neutrino mass degeneracy by introducing a Majorana type phase $\psi$. Next, we added a perturbation symmetric under the (23) subgroup of $S_3$. We chose the parameter $t$ for this perturbation in a way to eliminate the smaller ``solar neutrino" mass difference. The second perturbation with complex parameter $ve^{-i\phi}$, living in the (12) subspace, was added in a similar way with $\phi$ related to Dirac type phase. 
 
 In Sec III and IV, we carried out the diagonalization of the resultant complex symmetric matrix and determined the parameters of the model. We used the approximation in which we kept terms of the type $t/|\alpha|$, $v/t$ and $v/|\alpha|$. The ratio $v/t$ is effectively of first order and is less than 1/10 in magnitude. The solar neutrino mass difference came out to be proportional to the second order parameter $v$ and the atmospheric neutrino mass difference to the first order $t$ as well as second order $v$. Both mass differences were also dependent on the Majorana phase $\psi$ and  the Dirac phase $\phi$. For the mixing angles, the corrections had a piece proportional to the complex parameter $\epsilon$ which is effectively of the first order parameter $v/t$. 
 
Important motivations for this model were the non-zero value of $s_{13}$ and the Dirac CP phase. We found that the mixing angle $|s_{13}|$ was related to the deviation of $s_{23}$ from its tribimaximal value which we used as one of our experimental inputs. The predicted values for $|s_{13}|$ were close to the numbers from data (see Table \ref{eps2tab}). Our results for the CP-violating phase $\delta$ were well in agreement with the suggested value from global analysis of neutrino oscillation data \cite{fogli1} - \cite{fogli2} for inverse hierarchy and an order less for normal hierarchy (see Table \ref{deltatab}). The mixing angle $s_{12}$ was unchanged from its tribimaximal value by the perturbation in this model.

    The leptonic factor describing the amplitude for the neutrinoless double beta decay of a nucleus: $(A,Z)\rightarrow (A,Z+2) + 2e^-$ comes out to be,
\begin{equation}
|m_{ee}|\approx\frac{1}{3}|\alpha|\sqrt{5+4cos\psi}\left[1-\frac{v}{|\alpha|}\left(\frac{3sin\frac{\psi}{2}cos\phi-5cos\frac{\psi}{2}sin\phi}{5+4cos\psi}\right)\right].
\label{nubetadecayresult}
\end{equation}
    It may be noted that $|m_{ee}|$ depends also on the Dirac CP phase $\phi$ but this effect is negligible since it appears at second order. Some predictions for $|m_{ee}|$ at zeroth order for typical values of $\psi$ are listed in the following chart.

\begin{table}[htbp]
\begin{center}
\begin{tabular}{cccc}
  $\psi$: &  $\pi $,  & $\pi /2 $, & $\pi /4 $\\
$|m_{ee}|:$ & 0.033 eV,  & 0.074 eV, & 0.093 eV \\
\end{tabular}
\end{center}
 \label{meeatab}
\end{table} 
The current experimental bound on $|m_{ee}|$ is $<$ 0.36 eV  \cite{nulessbeta}. It can be seen that $|m_{ee}|$ gets larger for smaller values of the Majorana phase, $\psi$, for degenerate neutrino masses.
    
    We have not discussed the role of the Higgs sector in this paper but some technical details were discussed in \cite{jns}. We have also assumed that the diagonalization of the charged lepton mass matrix makes a negligible contribution to the lepton mixing matrix. Of course, at a later stage, the contributions of the charged leptons should be included at the expense of additional parameters or assumptions. Further results on this model will
be given elsewhere.

\section{acknowledgments}

The work of J. Schechter and M. N. Shahid was supported in part by the U.S. DOE under Contract No. DE-FG-02-85ER 40231. The work of R. Jora was supported by PN 09370102/2009. One of us (J.S.) would like to thank Prof. F. Sannino and the members of the CP$^3$ Origins group at Southern Denmark University for their warm hospitality and support during the Spring semester of 2012. MNS would like to thank Prof. A. Smirnov for hospitality and support at Abdus Salam International Centre for Theoretical Physics, Trieste where a part of this work was done.


\begin{thebibliography}{9}

\bibitem{jns}R. Jora, S. Nasri and J. Schechter, Int. J. Mod. Phys. A,
{\bf 21}, 5875 (2006).

\bibitem{cw}C.-Y. Chen and L. Wolfenstein, Phys. Rev. D {\bf 77},
093009 (2008).

\bibitem{jss2009}R. Jora, J. Schechter and M. Naeem Shahid, Phys. Rev. D ${\bf 80}$, 093007 (2009); Erratum D {\bf 82}, 079902 (2010).

\bibitem{jss2010}R. Jora, J. Schechter and M. Naeem Shahid, Phys. Rev. D ${\bf 82}$, 053006 (2010).

\bibitem{dgg1}S. Dev, Shivani Gupta and Radha Raman Gautam, Phys. Lett. B ${\bf 702}$, 28 (2011);  arXiv:1106.3873 [hep-ph].

\bibitem{dgg2} S. Dev, Radha Raman Gautam and Lal Singh, Phys. Lett. B ${\bf 708}$, 284 (2012); arXiv:1201.3755 [hep-ph].

\bibitem{w}L. Wolfenstein, Phys. Rev. D {\bf 18}, 958 (1978).

\bibitem{ps1}S. Pakvasa and H. Sugawara, Phys. Lett. B {\bf 73}, 61 (1978).

\bibitem{ps2}S. Pakvasa and H. Sugawara, Phys. Lett. B {\bf 82}, 105 (1979).

\bibitem{ps3}E. Derman and H.S.Tsao, Phys. Rev. D {\bf 20}, 1207 (1979). 

\bibitem{ps4}Y. Yamanaka, H. Sugawara and S. Pakvasa Phys. Rev. D {\bf 25}, 1895 (1982).

\bibitem{cfm}S.-L. Chen, M. Frigerio and E. Ma, hep-ph/0404084.

\bibitem{fty}M. Fukugita, M. Tanimoto and T. Yanagida, Phys. Rev. D {\bf 57}, 4429 (1998), hep-ph/9709388.

\bibitem{mr}E. Ma and G. Rajasekaran, Phys. Rev. D {\bf 64}, 113012 (2001), hep-ph/0106291.

\bibitem{xyz1}Z-z. Xing, D. Yang and S. Zhou, arXiv:1004.4234v2[hep-ph].
 
\bibitem{xyz2}Shun Zhou, Phys. Lett. B {\bf 704}, 291 (2011), arXiv:1106.4808[hep-ph].

\bibitem{dgr1}D. A. Dicus, S-F. Ge and W. W. Repko, arXiv:1004.3266[hep-ph]. 

\bibitem{dgr2}Shao-Feng Ge, Duane A. Dicus and Wayne W. Repko, Phys. Lett. B {\bf 702}, 220 (2011), arXiv:1104.0602[hep-ph].

\bibitem{dgr3}Shao-Feng Ge, Duane A. Dicus and Wayne W. Repko, Phys. Rev. Lett. {\bf 108}, 041801 (2012), arXiv:1108.0964[hep-ph].

\bibitem{dgr4}A. Yu. Smirnov,  arXiv:1210.4061[hep-ph].

\bibitem{fx}H. Fritzsch and Z.-Z.Xing, Phys. Lett. B {\bf 440}, 313 (1988), hep-ph/9808272.

\bibitem{HPS}P. F. Harrison, D. H. Perkins and W. G. Scott, Phys.
Lett. {\bf B530},79 (2002), hep-ph/0202074.

\bibitem{X}Z.-Z.Xing, Phys. Lett. B {\bf 533}, 85 (2002),
 hep-ph/020409.

\bibitem{HZ}X.G.He and A. Zee, Phys. Lett. B {\bf 560}, 87 (2003),
hep-ph/0204049.

\bibitem{HS} P. F. Harrison and W. G. Scott, hep-ph/0302025.

\bibitem{LV}C.I.Low and R.R.Volkas, Phys. Rev. D {\bf 68},
033007 (2003), hep-ph/0305243.

\bibitem{Z}A.Zee, Phys. Rev. D {\bf 68}, 093002 (2003),
hep-ph/0307323.

\bibitem{BHS} J.D. Bjorken, P. F. Harrison and W. G. Scott,
hep-ph/0511201.

\bibitem{MNY} R.~N.~Mohapatra, S.~Nasri and H.~B.~Yu,
  arXiv:hep-ph/0605020.
  
\bibitem{king1}S.~F.~King, Nucl.\ Phys.\ B {\bf 576}, 85 (2000).
   
\bibitem{king2}S.~F.~King and N.~N.~Singh, Nucl.\ Phys.\ B {\bf 591}, 3 (2000).

\bibitem{mamain} E. Ma, Phys. Rev. D {\bf 70} 091301 (2004).

\bibitem{hvvm}M. Hirsch, A. Velanova del Morel, J.W.F. Valle and 
E. Ma, Phys. Rev. D {\bf 72}, (031901) (2005).

\bibitem{mmp}A. Montdragon, M. Montdragon and E. Peinado, J.Phys.
 {\bf A41}, 304035 (2008)[ArXiv:0712.1799]; see
 also arXiv:0805.3507.

\bibitem{mutau1}T. Fukuyama and H. Nishiura, hep-ph/9702253.

\bibitem{mutau2}R. N. Mohapatra and S. Nussinov, Phys. Rev. {\bf D 60}, 013002
(1999). 

\bibitem{mutau3}E. Ma and M. Raidal, Phys. Rev. Lett. {\bf 87}, 011802
(2001).

\bibitem{mutau4}C. S. Lam, hep-ph/0104116.

\bibitem{mutau5}T. Kitabayashi and M. Yasue,
Phys.Rev. {\bf D67} 015006 (2003).

\bibitem{mutau6}W. Grimus and L. Lavoura,
hep-ph/0305046; 0309050.

\bibitem{mutau7}Y. Koide, Phys.Rev. {\bf D69}, 093001
(2004).

\bibitem{mutau8}Y. H. Ahn, Sin Kyu Kang, C. S. Kim and Jake Lee, hep-ph/0602160.

\bibitem{mutau9}A. Ghosal, hep-ph/0304090.

\bibitem{mutau10}W. Grimus and L. Lavoura, Phys.\ Lett.\ B {\bf 572}, 189 (2003).

\bibitem{mutau10}W.~Grimus and L.~Lavoura, J.\ Phys.\ G {\bf 30}, 73 (2004).

\bibitem{moh1}W. Grimus, A. S.Joshipura, S. Kaneko, L.
Lavoura, H. Sawanaka and M. Tanimoto, hep-ph/0408123.

\bibitem{moh2}R. N. Mohapatra,
JHEP, {\bf 0410}, 027 (2004).

\bibitem{moh3}A. de Gouvea, Phys.Rev. {\bf D69},
093007 (2004).

\bibitem{moh4}R. N. Mohapatra and W. Rodejohann, Phys. Rev. {\bf D
72}, 053001 (2005).

\bibitem{moh5}T. Kitabayashi and M. Yasue, Phys. Lett,. {\bf B
621}, 133 (2005). 

\bibitem{moh6}R.~N.~Mohapatra and S.~Nasri, Phys.\ Rev.\ D {\bf
71}, 033001 (2005).

\bibitem{moh7}R.~N.~Mohapatra, S.~Nasri and H.~B.~Yu, Phys.\
Lett.\ B {\bf 615}, 231 (2005).

\bibitem{Gutmutau1}K. Matsuda and H. Nishiura,
 Phys.\ Rev.\ D {\bf 73}, 013008 (2006).
 
\bibitem{Gutmutau2}A. Joshipura, hep-ph/0512252.

\bibitem{Gutmutau3}R. N. Mohapatra, S. Nasri and H.~B.~Yu, Phys. Lett. {\bf B 636}, 114
(2006).

\bibitem{ks}T. Kaneko, H. Sugawara, Phys. Lett. B ${\bf 697}$, 329 (2011) .

\bibitem{bkns1}A. Bottino, C. W. Kim, H. Nishiura and W. K. Sze, Phys. Rev. D {\bf 34}, 862 (1986).

\bibitem{bkns2}R. Johnson, S. Ranfone and J. Schechter, Phys. Rev. D. {\bf 35}, 282 (1987).

\bibitem{superK}Super Kamiokande collaboration, S. Fukuda et al,
Phys. Lett. B {\bf 539}, 179 (2002), hep-ex/0205075.

\bibitem{kamland}KamLAND collaboration, K. Eguchi et al, Phys. Rev.
Lett. {\bf 90}, 021802 (2003).

\bibitem{sno}SNO collaboration, Q. R. Ahmad et al,nucl-ex/
0309004.

\bibitem{k2k}K2K collaboration, M. H. Ahn et al, Phys. Rev. Lett. {\bf
90}, 041801 (2003).

\bibitem{gall}GALLEX Collaboration, W. Hampel et al, Phys. Lett. B
{\bf 447}, 127 (1999).

\bibitem{sage}SAGE Collaboration, J. N. Abdurashitov et al,
Phys. Rev. C {\bf 60}, 055801 (1999).

\bibitem{chooz}CHOOZ Collaboration, M. Apollonio et al,
Eur. Phys. J. C {\bf 27}, 331 (2003), hep-ex/0301017.

\bibitem{minos}MINOS Collaboration, Phys. Rev. D {\bf 73}, 072002
 (2005), hep-ex/0512036.

\bibitem{A}K. Nakamura {\it et al} (Particle Data Group) 2010, J. Phys. G: Nucl. Part. Phys. {\bf 37} 075021.

\bibitem{cosmobound1}D. N. Spergel {\it et al}, Astrophys. J. Suppl. {\bf 148}, 175 (2003).

\bibitem{cosmobound2}S. Hannestad, JCAP {\bf 0305}, 004 (2003).

\bibitem{trimax}Sandip Pakvasa, Werner Rodejohann, Thomas J. Weiler, Phys. Rev. Lett. {\bf 100}, 111801 (2008), arXiv:0711.0052 [hep-ph].

\bibitem{fogli1}G.L. Fogli {\it et al}., arXiv:1205.5254..

\bibitem{fogli2}D. V. Forero, M. Tórtola and J. W. F. Valle, arXiv:1205.4018.

\bibitem{nulessbeta}J. J. Gomez-Cadenas, J. Martin-Albo, M. Mezzetto, F. Monrabal and M. Sorel, Riv. Nuovo Cim. {\bf 35}, 29 (2012), arXiv:1109.5515 [hep-ex].

\end{thebibliography}
\end{document}